\documentclass[preprint,showpacs,floatfix]{revtex4}
\usepackage[dvips]{graphicx}
\begin{document}
\title{Origin of $2_1^+$ Excitation Energy Dependence \\on Valence Nucleon Numbers}
\author{Eunja \surname{Ha}}
\author{Dongwoo \surname{Cha}}
\email{dcha@inha.ac.kr}
\thanks{Fax: +82-32-866-2452}
\affiliation{Department of Physics, Inha University, Incheon
402-751, Korea}
\date{March 13, 2007}

\begin{abstract}
It has been shown recently that a simple formula in terms of the
valence nucleon numbers and the mass number can describe the
essential trends of excitation energies of the first $2^+$ states in
even-even nuclei. By evaluating the first order energy shift due to
the zero-range residual interaction, we find that the factor which
reflects the effective particle number participating in the
interaction from the Fermi orbit governs the main dependence of the
first $2^+$ excitation energy on the valence nucleon numbers.
\end{abstract}


\maketitle

Valence nucleon numbers $N_p$ and $N_n$ have been extensively
adopted to parameterize various nuclear properties
phenomenologically. The valence proton (neutron) number $N_p\,\,
(N_n )$ is defined as the number of proton (neutron) particles below
the mid-shell or the number of proton (neutron) holes past the
mid-shell within the given major shell. Hamamoto was the first to
show that the square root of the ratio of the measured and the
single-particle $B(E2)$ values, $ \left[ B(E2)_{\rm exp} /
B(E2)_{\rm sp} \right]^{1/2} $, is roughly proportional to the
product $N_p N_n $ \cite{Hamamoto}. Casten extended the idea and
suggested the $N_p N_n$ scheme where he showed that if we
parameterize the collective variables or operators in even-even
nuclei in terms of the product $N_p N_n$ then we get a substantial
reduction in the number of parameters without serious loss of
accuracy \cite{Casten,Casten-a}. It was also demonstrated that the
$N_p N_n$ scheme could be applied not only to even-even nuclei but
also to odd-$A$ and odd-odd nuclei \cite{Zhao}.

\begin{figure}[b]
\centering
\includegraphics[width=8.0cm,angle=0]{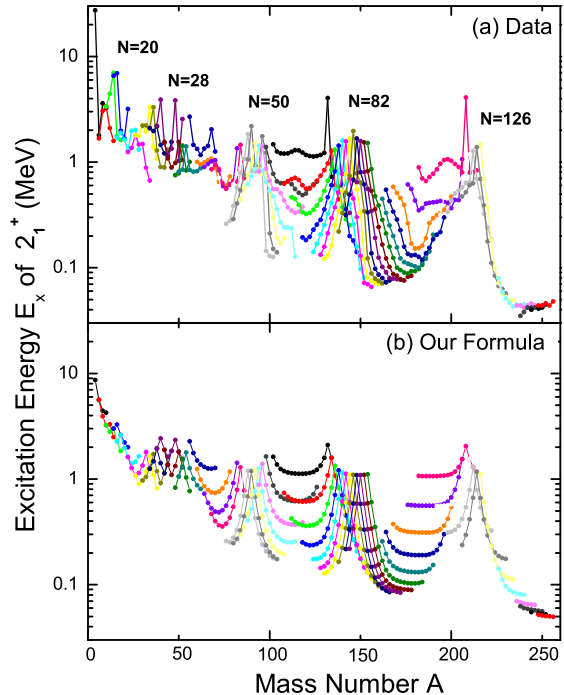}
\caption{Excitation energies of the first $2^+$ states in even-even
nuclei. The solid lines represent isotopic chains. Part (a) shows
the data quoted from Ref. 6, and part (b) shows the results obtained
by Eq.\,(\ref{E}).} \label{fig-1}
\end{figure}

Recently, we proposed a simple empirical formula in terms of the
valence nucleon numbers $N_p$ and $N_n$, and the mass number $A$ for
the excitation energies $E_x (2_1^+ )$ of the first $2^+$ states in
even-even nuclei \cite{Ha}. The formula is given by
\begin{equation} \label{E}
E_x = \alpha A^{-\gamma} + \beta \left[ \exp ( - \lambda N_p ) +
\exp ( - \lambda N_n ) \right]
\end{equation}
where the parameters $\alpha$, $\beta$, $\gamma$, and $\lambda$ are
fitted from the data. From Fig.\,\ref{fig-1}, which is reproduced
from Ref. 5, we find that the essential trends of the excitation
energies $E_x (2_1^+ )$ are well reproduced throughout the periodic
table. The data displayed in part (a) of Fig.\,\ref{fig-1} is quoted
from the recent compilation by Raman {\it et al}. \cite{Raman} and
the results shown in part (b) of Fig.\,\ref{fig-1} were obtained by
applying Eq.\,(\ref{E}) where $\alpha = 34.9\, {\rm MeV}$, $\beta =
1.00 \, {\rm MeV}$, $\gamma = 1.19$, and $\lambda = 0.36$ \cite{Ha}.

In this brief report, we want to examine what makes such an odd
dependence of the excitation energies $E_x (2_1^+ )$ as given by
Eq.\,(\ref{E}) on the valence nucleon numbers $N_p$ and $N_n$. Since
we are dealing with the systematics observed throughout the whole
periodic table instead of inspecting the detailed structure of a
single nucleus, we want to keep the physics involved as simple as
possible. For that purpose, we first find the difference between the
energy of the lowest two quasi-particle state and the excitation
energy of the first $2^+$ state. Then we compare it with the first
order energy shift due to the residual interaction. Here, we obtain
the former energy difference from the measured quantities and the
latter energy shift we evaluate by using simple perturbation theory.

Within the standard BCS theory \cite{Bardeen}, the single
quasi-particle energy $E_k$ of the $k$-th orbit is given by
\begin{equation} \label{E-k}
E_k = \sqrt{(\epsilon_k - \mu )^2 + \Delta^2}
\end{equation}
where $\epsilon_k$, $\mu$, and $\Delta$ are the single particle
energy, the Fermi energy, and the gap energy, respectively. We can
obtain the gap energy $\Delta$ empirically from the binding energies
of the adjacent nuclei. Let $B(Z,N)$ be the negative binding energy
of a nucleus whose atomic number and neutron number are $Z$ and $N$,
respectively. Then the proton gap energy $\Delta_p$ and the neutron
gap energy $\Delta_n$ can be approximated by the following three
point expressions \cite{Kaneko}:
\begin{eqnarray}
\Delta_p \approx {1 \over 2} \left|  B(Z+1,N)  - 2 B(Z,N) +B(Z-1,N)
\right|,\label{D-p}
\\
\Delta_n \approx {1 \over 2} \left|  B(Z,N+1)  - 2 B(Z,N) +B(Z,N-1)
\right|.\label{D-n}
\end{eqnarray}

\begin{figure}[t]
\centering
\includegraphics[width=8.0cm,angle=0]{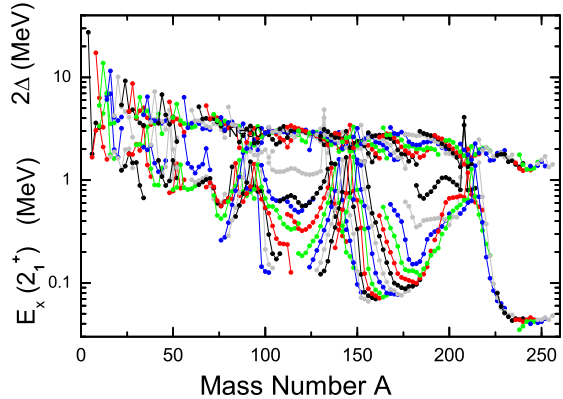}
\caption{Upper part shows twice the gap energy $2 \Delta$, which is
the lesser of $2 \Delta_p$ (Eq.\,(\ref{D-p})) and $2 \Delta_n$
(Eq.\,(\ref{D-n})). Binding energies are taken from Ref. 10. Lower
part shows the measured excitation energies of the first $2^+$
states (the same as those shown in Fig.\,\ref{fig-1}(a)).}
\label{fig-2}
\end{figure}

Let us write the excitation energy $E_x (2_1^+ )$ of the first $2^+$
state by simple perturbation theory as
\begin{equation}\label{E-x}
E_x (2_1^+ ) = E^{(0)} + E^{(1)} + \cdots .
\end{equation}
Then the unperturbed energy $E^{(0)}$ is equal to the energy of the
lowest two quasi-particle state, $E_k + E_{k'}$ where $k$ and $k'$
indicate the orbits which belong to the lowest two quasi-particle
state. Also the first order energy shift $E^{(1)}$ is given by
\begin{equation}\label{E-1}
E^{(1)} = < j_k j_{k'}|{\hat V}|j_k j_{k'}>_{J^\pi = 2^+}.
\end{equation}
which is the expectation value of the residual interaction $\hat V$
on the lowest two quasi-particle state which is coupled to the total
angular momentum $J^\pi =2^+$.

From Eq.\,({\ref{E-k}}), we have that $E_k \approx \Delta$ for the
single particle orbits $k$ near the Fermi level. Therefore, we take
twice the gap energy $2 \Delta$ in place of the lowest two
quasi-particle energy $E_k + E_{k'}$\cite{Satula}. At the upper part
of Fig.\,\ref{fig-2}, twice the gap energy, $2 \Delta$, is shown for
even-even nuclei. Here $2 \Delta$ is determined by the lesser of $2
\Delta_p$ and $2 \Delta_n$ which is estimated empirically either by
Eq.\,(\ref{D-p}) or by Eq.\,(\ref{D-n}) \cite{Audi}. We can observe
that all of the $2 \Delta$ are located higher in energy than the
measured $E_x (2_1^+ )$ data, which is shown at the lower part of
Fig.\,\ref{fig-2}. We, therefore, want to compare the difference
between $2 \Delta$ at the upper part and $E_x (2_1^+ )$ at the lower
part of Fig.\,\ref{fig-2} with the first order energy shift
$E^{(1)}$ which will be given shortly.

In order to estimate $E^{(1)}$, we take only the central part of the
residual interaction in the zero-range approximation,
\begin{equation}\label{V}
{\hat V} (1,2) = - v_0 \delta^3 ( {\vec r}_1 - {\vec r}_2 ),
\end{equation}
where $v_0$, which is the strength of the zero-range interaction,
will be treated as the only model parameter in our calculations.
Analytic expressions for the matrix elements of the zero-range
interactions are readily available from the literature \cite{Speth}.
The expectation value appearing in Eq.\,(\ref{E-1}) is given by
\begin{widetext}
\begin{eqnarray}
E^{(1)} &=& < j_k j_{k'} |{\hat V}|j_k j_{k'}>_{J^\pi = 2^+}
\nonumber \\
&=& -v_0 \frac{<R^4 >}{40 \pi} \frac{(2j_k +1)(2j_{k'}
+1)}{(1+\delta_{j_k j_{k'}})} \Big[ <j_k \frac{1}{2} j_{k'}
-\frac{1}{2} |J0>^2
\nonumber \\
&~&~~ \times \Big\{ [ 1+(-)^{j_k + j_{k'}} ] ( u_{j_k}^2
u_{j_{k'}}^2 + v_{j_k}^2 v_{j_{k'}}^2 ) + [2-(-1)^{j_k + j_{k'}} ]
(u_{j_k} v_{j_{k'}} + v_{j_k} u_{j_{k'}} )^2 \Big\}
\nonumber \\
&~&~~~~~~+ (-)^{j_k + j_{k'}} <j_k \frac{1}{2} j_{k'} \frac{1}{2}
|J1>^2 (u_{j_k} v_{j_{k'}} + (-)^{j_k + j_{k'}} v_{j_k} u_{j_{k'}}
)^2 \Big], \label{delta-V}
\end{eqnarray}
\end{widetext}
where $<R^4 >$, $v_{j_k}^2$, and $u_{j_k}^2$ are the integral of the
radial wave functions
\begin{equation}\label{R-4}
<R^4 > = \int_0^{\infty} dr \, r^2 \left[ R_{j_k} (r) R_{j_{k'}}
\right]^2,
\end{equation}
the occupation probability of the $k$-th orbit, and $u_{j_k}^2 = 1 -
v_{j_k}^2$, respectively.

We evaluate the matrix element in Eq.\,(\ref{delta-V}) by using the
following procedure. First, the single-particle wave functions and
energies are generated by diagonalizing the Woods-Saxon (WS)
potential
\begin{equation}\label{WS}
V_{WS} (r) = - {{V_0} \over {1+ \exp \left[ {{r-R_0} \over a}
\right] }}
\end{equation}
where for the WS parameters we use $V_0 = 50 \,{\rm MeV}$, $R_0 =
1.27 \, {\rm fm}$, and $a=0.67\,{\rm fm}$. Then the radial integral
$<R^4>$ of Eq.\,(\ref{R-4}) is evaluated using the WS wave
functions. Also the occupation probabilities $v_{j_k}^2$ in
Eq.\,(\ref{delta-V}) are determined by solving the BCS equations
with the simple monopole pairing interaction\cite{Lane}
\begin{equation}\label{pairing}
<(j_1 j_2 )J|v_{\rm pair} | (j_3 j_4 )J> = - {1 \over 2} G
\delta_{J,0} \sqrt{(2 j_1 +1 )(2 j_3 +1)}
\end{equation}
where the strength $G$ of the pairing interaction is adjusted to
reproduce the measured gap energy by Eq.\,(\ref{D-p}) or
Eq.\,(\ref{D-n}) for each nucleus in our calculations.

\begin{figure}[t]
\centering
\includegraphics[width=8.0cm,angle=0]{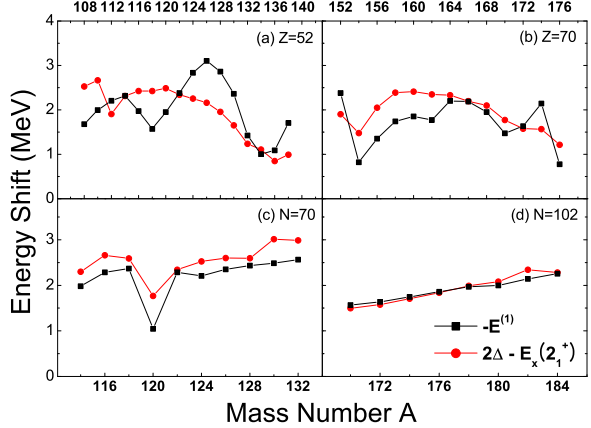}
\caption{Calculated results (solid squares) for the energy shift
$-E^{(1)}$ are compared with the measured differences (solid
circles) $2 \Delta - E_x (2_1^+ )$ for the following four cases: (a)
$Z=52$ isotopic chain, (b) $Z=70$ isotopic chain, (c) $N=70$
isotonic chain, and (d) $N=102$ isotonic chain.} \label{fig-3}
\end{figure}

In Fig.\,\ref{fig-3}, we compare our calculated results (solid
squares) for $-E^{(1)}$ with the measured differences (solid
circles) $2 \Delta - E_x (2_1^+ )$ for the following four cases: (a)
$Z=52$ isotopic chain, (b) $Z=70$ isotopic chain, (c) $N=70$
isotonic chain, and (d) $N=102$ isotonic chain. In our calculations,
the strength $v_0$ of the zero-range interaction in Eq.\,(\ref{V})
is adjusted so that the calculated $-E^{(1)}$ coincides exactly with
the measured difference $2 \Delta - E_x (2_1^+ )$ for the heaviest
lead isotope $_{~82}^{184} {\rm Pb}$ in Fig.\,\ref{fig-3}(d). The
resulting adopted value of $v_0$ in all of our subsequent
calculations for $E^{(1)}$ is $v_0 = 1,700\,{\rm MeV \, fm^3}$. By
observing the results displayed in Fig.\,\ref{fig-3}, we find that
the first order energy shift $-E^{(1)}$ calculated by such a simple
model like perturbation theory describes the measured difference $2
\Delta - E_x (2_1^+ )$ unbelievably well. Thus, this results enable
us to examine $E^{(1)}$ in place of $E_x (2_1^+ )$ to find the
origin of its dependence on the valence nucleon numbers $N_p$ and
$N_n$.

If the total angular momentum $j_f$ of the Fermi orbit is larger
than $1/2$, which holds practically in most cases of our
calculations, then both of the two quasi-particles which form the
lowest two quasi-particle state belong to the Fermi orbit. Therefore
$j_k$ and $j_{k'}$ in Eq.\,(\ref{delta-V}) can be replaced by $j_f$.
This, in turn, reduces the expression for the energy shift $E^{(1)}$
in Eq.\,(\ref{delta-V}) to a very simple form which is written as
\begin{eqnarray}
E^{(1)} &=& < j_f j_f |{\hat V}|j_f j_f>_{J^\pi = 2^+}
\nonumber \\
&=& \left[- \frac{3V_0}{20\pi} <R^4 > <j_f {1 \over 2} j_f -{1 \over
2}|2,0>^2 \right] \Big[ (2j_f +1)^2 u_{j_f}^2 v_{j_f}^2
\Big]\label{reduced-V}
\end{eqnarray}
where we use the known value of the Clebsch-Gordan coefficient
\begin{equation}\label{clebsh}
<j {1 \over 2} j {1 \over 2} |J,1>=0
\end{equation}
which holds identically for any even $J$ and half integer values of
$j$.

\begin{figure}[t]
\centering
\includegraphics[width=8.0cm,angle=0]{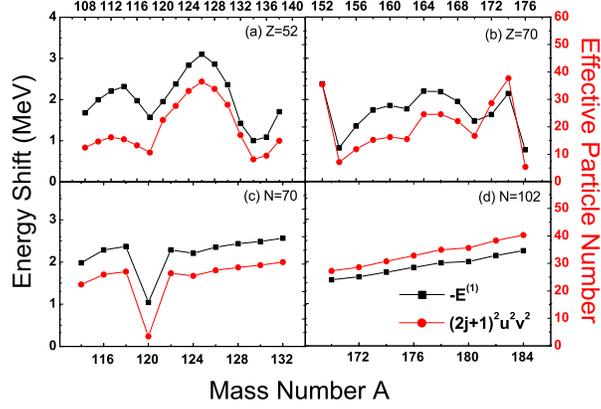}%
\caption{Calculated results (solid squares) for the energy shift
$-E^{(1)}$ are compared with the effective particle number $(2j_f
+1)^2 u_{j_f}^2 v_{j_f}^2$ for the same four cases as those in
Fig.\,\ref{fig-3}. Note that the scale for the energy shift is shown
at the far left axis while that for the effective particle number is
shown at the far right axis.} \label{fig-4}
\end{figure}

The expression for the first order energy shift $E^{(1)}$ can be
divided into two major factors as shown by the large square brackets
in Eq.\,(\ref{reduced-V}). One is the radial integral $<R^4
>$ times the Clebsch-Gordan coefficient squared and the other is
$(2j_f +1)^2 u_{j_f}^2 v_{j_f}^2$ which can be interpreted to
reflect the number of effective particles that participate in the
interaction from the Fermi orbit. Out of the two factors, we expect
that, the former is kept more or less the same over different
isotopes and isotones. Therefore, the variation in values of
$E^{(1)}$ follows that of the effective particle number $(2j_f +1)^2
u_{j_f}^2 v_{j_f}^2$. In order to check this point, we compare the
negative of the first order energy shift $-E^{(1)}$ (solid squares)
with the effective particle number $(2j_f +1)^2 u_{j_f}^2 v_{j_f}^2$
(solid circles) in Fig.\,\ref{fig-4}  for the same four cases as
those in Fig.\,\ref{fig-3}. Note that the scale for the energy shift
is shown at the far left axis while that for the effective particle
number is shown at the far right axis in Fig.\,\ref{fig-4}. We can
confirm undoubtedly from Fig.\,\ref{fig-4} that the behavior of
$E^{(1)}$ follows that of the effective particle number quite
closely.

\begin{figure}[t]
\centering
\includegraphics[width=8.0cm,angle=0]{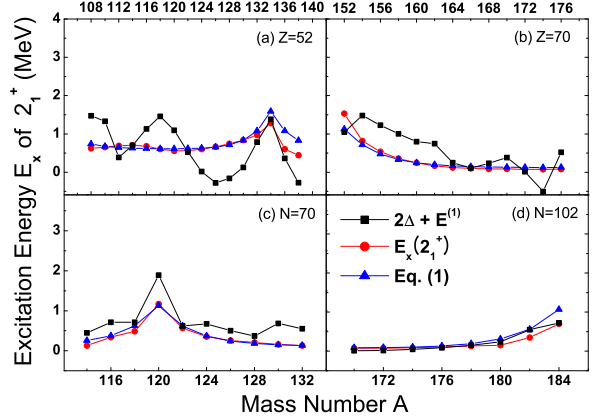}%
\caption{Measured excitation energies (solid circles) $E_x (2_1^+ )$
of the first $2^+$ states for the same four cases as those in
Fig.\,\ref{fig-3} are compared with the one calculated by the
perturbation theory (solid squares), namely $2 \Delta + E^{(1)}$,
and with the one obtained from the empirical formula (solid
triangles) given by Eq.\,(\ref{E}). Measured excitation energies
$E_x (2_1^+)$ are quoted from Ref.\,6.} \label{fig-5}
\end{figure}

Finally, we compare the measured excitation energies (solid circles)
$E_x (2_1^+ )$ of the first $2^+$ states in Fig.\,\ref{fig-5} for
the same four cases as those in Fig.\,\ref{fig-3} with the one
calculated by perturbation theory (solid squares), namely $2 \Delta
+ E^{(1)}$, and with the one obtained from the empirical formula
(solid triangles) given by Eq.\,(\ref{E}). By observing these plots,
we can ascertain, first of all, that the empirical formula,
Eq.\,(\ref{E}), expressed in terms of the valence nucleon numbers
describes the measured excitation energies $E_x(2_1^+)$ quite well,
even quantitatively to some extent. Furthermore, we can also confirm
that simple perturbation theory can be employed in examining the
possible dependence of $E_x(2_1^+)$ on the valence nucleon numbers
at least qualitatively.

In summary, we have examined the recently proposed empirical
formula, Eq.\,(\ref{E}), for the possible origin of its dependence
on valence nucleon numbers. Recently, it has been shown that
Eq.\,(\ref{E}), which depends on the mass number $A$ and the valence
nucleon numbers $N_p$ and $N_n$ in a very simple fashion, can
describe the essential trends of excitation energies $E_x(2_1^+)$ of
the first $2^+$ states in even-even nuclei throughout the whole
periodic table \cite{Ha}. In order to find out what makes such a
dependence of $E_x(2_1^+)$ on $N_p$ and $N_n$, we calculated the
first order energy shift $E^{(1)}$ resulting from the zero-range
residual interaction by using simple perturbation theory. Then the
shif was compared with the difference between the energy of the
lowest two quasi-particle state and the measured excitation energy
$E_x(2_1^+)$. The lowest two quasi-particle energy was approximated
by twice the gap energy determined from the binding energies of the
adjacent nuclei through the three point expression given by
Eq.\,(\ref{D-p}) or (\ref{D-n}). In our calculations, the strength
$v_0$ of the zero-range residual interaction given by
Eq.\,({\ref{V}) was the only free parameter. It was kept fixed in
all of our calculations at $v_0 = 1,700\,{\rm MeV fm^3}$ for which
the calculated $-E^{(1)}$ coincided exactly with the measured
difference $2\Delta - E_x (2_1^+)$ for the heaviest lead isotope
$_{~82}^{184}{\rm Pb}$. We found that the variation in values of
$E^{(1)}$ followed that of the factor $(2j_f +1 )^2 u_{j_f}^2
v_{j_f}^2$ which can be interpreted as the effective particle number
participating in the interaction from the Fermi orbit. Therefore, we
concluded that the effective particle number governed the main
dependence of the first $2^+$ excitation energy on the valence
nucleon numbers as given by Eq.\,(\ref{E}).

\begin{acknowledgments}
We are grateful to Jin-Hee Yoon for useful discussions. This work
was supported by an Inha University research grant.
\end{acknowledgments}

\end{document}